# Precision microfluidic control of neuronal ensembles in cultured cortical networks

*Hakuba Murota, Hideaki Yamamoto*, Nobuaki Monma, Shigeo Sato, and Ayumi Hirano-Iwata*

H. Murota, H. Yamamoto, N. Monma, S. Sato, A. Hirano-Iwata
Research Institute of Electrical Communication, Tohoku University, Sendai 980-8577, Japan
E-mail: hideaki.yamamoto.e3@tohoku.ac.jp

H. Murota, H. Yamamoto, N. Momna, S. Sato, A. Hirano-Iwata
Graduate School of Engineering, Tohoku University, Sendai 980-8579, Japan

H. Yamamoto, A. Hirano-Iwata
Advanced Institute for Materials Science, Tohoku University, Sendai 980-8577, Japan

**Abstract**

In vitro neuronal culture is an important research platform in cellular and network neuroscience. However, neurons cultured on a homogeneous scaffold form dense, randomly connected networks and display excessively synchronized activity; this phenomenon has limited their applications in network-level studies, such as studies of neuronal ensembles, or coordinated activity by a group of neurons. Herein, we develop polydimethylsiloxane-based microfluidic devices to create small neuronal networks exhibiting a hierarchically modular structure resembling the connectivity observed in the mammalian cortex. The strength of intermodular coupling was manipulated by varying the width and height of the microchannels that connect the modules. Using fluorescent calcium imaging, we observe that the spontaneous activity in networks with smaller microchannels (2.2-5.5 $\mu m^2$) had lower synchrony and exhibit a threefold variety of neuronal ensembles. Optogenetic stimulation demonstrates that a reduction in intermodular coupling enriches evoked neuronal activity patterns and that repeated stimulation induces plasticity in neuronal ensembles in these networks. These findings suggest that cell engineering technologies based on microfluidic devices enable in vitro reconstruction of the intricate dynamics of neuronal ensembles, thus providing a robust platform for studying neuronal ensembles in a well-defined physicochemical environment.

## 1. Introduction

Neuronal ensembles are coordinated activities of neurons that occur in multiple regions of the mammalian brain. For example, in the rodent visual cortex, visual stimulation evokes a set of neuronal ensembles that are present in spontaneous neural activity.[1,2] Such properties of neuronal ensembles have also been reported in the auditory cortex.[3] The importance of neuronal ensembles has also been highlighted in the somatosensory cortex, where it has been shown that activation of specific ensembles related to pain triggers associated behaviors in mice.[4] The statistics of neuronal ensembles shift over time, and their plasticity in the hippocampus has been associated with memory formation.[5] These findings support the hypothesis that neuronal ensembles are endogenous building blocks of neural circuits[6] and that the understanding of their function is important for elucidating the mechanisms of neural information processing.

In vivo animal experiments continue to be the major platform for studying neuronal network functions, including the properties of neuronal ensembles. As an alternative approach, in vitro neuronal culture offers unparalleled access for imaging, manipulation, and control of living neuronal networks,[7] thus providing a unique tool for studying network functions. However, neurons in homogeneous cultures develop to form a randomly connected network that exhibits bursting activity that is synchronized across a large population.[8,9] To examine neuronal ensemble functions in vitro, network development must be guided to allow cultured neurons to organize into a network that resembles the intricate wiring structure of in vivo networks so that excessively synchronized activity is suppressed.

Cell engineering technologies using micropatterned substrates,[10,11] microfluidic devices[12], or engineered scaffolds[13,14] have been developed for this purpose. Notably, microfluidic devices are an outstanding method for patterning neurons due to their ability to precisely define structures and their ease of handling. Most microfluidic devices that are used in neuroscience research have common structures such as reservoirs and microchannels, which confine the locations of cell bodies and neurite outgrowth, respectively.[12] Importantly, the number of neurites that interconnect neighboring reservoirs (i.e., the strength of their functional coupling) can be precisely controlled by the size of the microchannels.[15-17]

In this study, we fabricated and characterized microfluidic devices with diverse microchannels to constrain intermodular coupling and suppress excessive network-wide synchrony for neuronal ensemble analysis in cultured neuronal networks. The reservoir and microchannels are designed such that the overall network bears a hierarchically modular structure, which is a canonical structure that is evolutionarily preserved in the nervous systems of animals.[18,19] The cross-sectional area of the microchannels decreased to 2.2 μm$^2$, which is one to two orders of magnitude smaller than that of generally used microchannels.[20-24] Neurons that adhered inside of the module (reservoir) extended neurites into the microchannels even when their size was reduced to 2.2 μm$^2$. Recordings of spontaneous neural activity by fluorescence calcium imaging showed that networks grown in microfluidic devices with microchannels below 5.5 μm$^2$ exhibit multiple neuronal ensembles, which could be manipulated by repetitive optogenetic stimulation. Taken together, our device allows for the reconstitution of neuronal networks and serves as a novel model system for investigating the plasticity and stability of neuronal ensembles.

## 2. Results

### 2.1. Engineered neuronal networks in small-channel microfluidic devices

Polydimethylsiloxane (PDMS) microfluidic devices for patterning primary rat cortical neurons were fabricated by using the replica molding method following the method described in a previous study (Fig. 1A-C).[20] Each network was composed of 16 modules that were connected in a hierarchically modular arrangement (Fig. 1D-E).[25] Multiple neurons grew and formed dense interconnections within each module, and neighboring modules were connected with microchannels, wherein axons and dendrites grew to functionally couple the neurons in the modules (Fig. 1F).

The master mold for structuring the PDMS was fabricated via photolithography to pattern two layers of SU-8 photoresist. The width and height of the microchannels were varied to manipulate the number of neurites that interconnected neighboring modules (Fig. 1A). Microchannels with widths of 2.3 ± 0.3 and 3.6 ± 0.4 μm (mean ± standard deviation (SD)) were fabricated by using photomasks with 1 and 2 μm-wide metal patterns, respectively (Fig. 1G). The microchannel height was controlled by adjusting the viscosity of the first-layer photoresist by adding γ-butyrolactone (GBL) to SU-8 3005.[26] We found that with a decreasing fraction of SU-8 3005 in the SU-8/GBL mixture, the thickness of the first layer decreased accordingly from 2.53 μm to 0.40 μm (Fig. 1H).

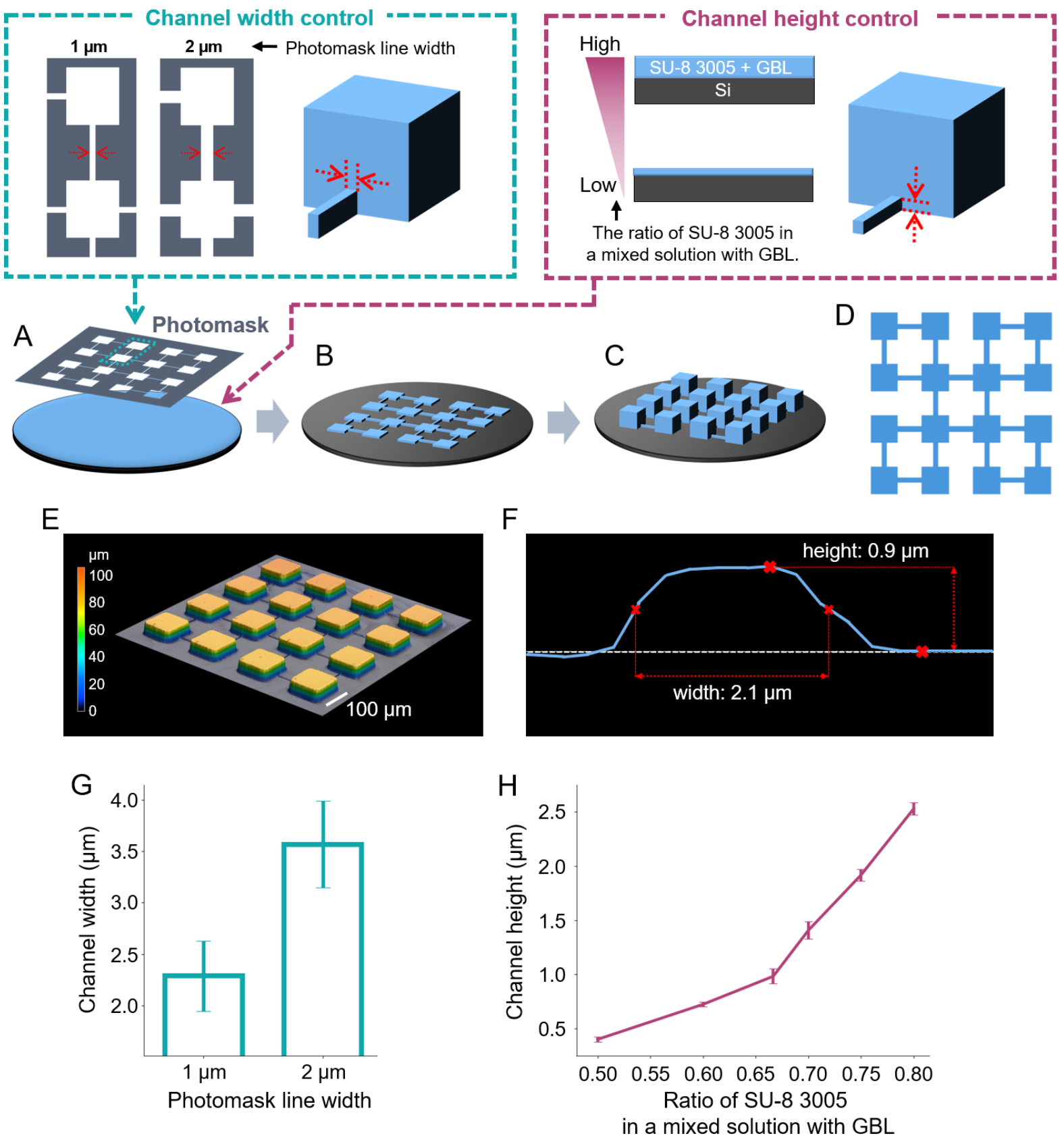

Figure 1. Microchannel design of molds for the fabrication of microfluidic devices by changing photolithography conditions. (A) Spin-coating of SU-8 3005 and GBL on a silicon wafer and exposure through a photomask. We prepared photomasks with two different line widths (1 and 2 μm) to change the width of the microchannel (top left) and adjusted the ratio of SU-8 3005 to GBL to achieve a change in the microchannel height (top right). (B) Development after exposure. (C) The obtained sample was then spin-coated with SU-8 3050, and photolithography was performed. (D) Schematic illustration of the micropattern. (E) 3D reconstruction of the master mold. (F) Line profile of cross-section of the microchannel. The height and width of the microchannels were defined as the distance from the reference plane to the highest point and the full width at half maximum, respectively. (G) Relationship between the photomask line width and the microchannel width at half of the height. The bars represent the mean, and error bars are the SDs. (H) Relationship between the ratio of SU-8 3005 in a mixed solution with GBL and the microchannel height. The line represents the mean, and error bars are the SDs.

The fabricated microfluidic devices were attached to poly-D-lysine (PDL)-coated coverslips, and cortical neurons were seeded and cultured in the devices for 14–15 days (Figs. 2A-B). Microfluidic devices featuring five different cross-sectional areas were prepared: 2.2, 3.4, 3.7, 5.5, and 44.5 μm$^2$. Due to the fact that the cross-sectional areas of the microchannels used for neuronal patterning are typically several tens of μm$^2$[20-24], the devices bearing microchannels with the first four areas are referred to as *small-channel* devices, whereas the device with microchannels with the last area is referred to as a *conventional-channel* device (Fig. 2C).

To verify the development of cultured neurons, particularly whether neurons extend neurites into microchannels, we analyzed neurons growing in small-channel devices (2.2 μm$^2$) by using confocal microscopy. We found that neurons within the module formed a three-dimensional aggregate and that neurites extending out from the aggregate entered a microchannel even when the channel size was reduced to 2.2 μm$^2$ (Figs. 2D-E). Due to the fact that the neurites formed bundles in the microchannel, the quantification of the exact number of neurites was technically challenging; however, the presence of multiple neurites was observed in all of the microchannels ($n$ = 10 channels; Fig. S1). Neuronal somas were not observed inside of the microchannels, in agreement with a previous study reporting that a 1.5 μm-high channel prevented neuronal soma migration even when the channel width was as wide as 15 μm.[27] In summary, neuronal networks with modular topology were successfully created, thus featuring microchannels with cross-sectional areas more than ten times smaller than those of conventional networks.

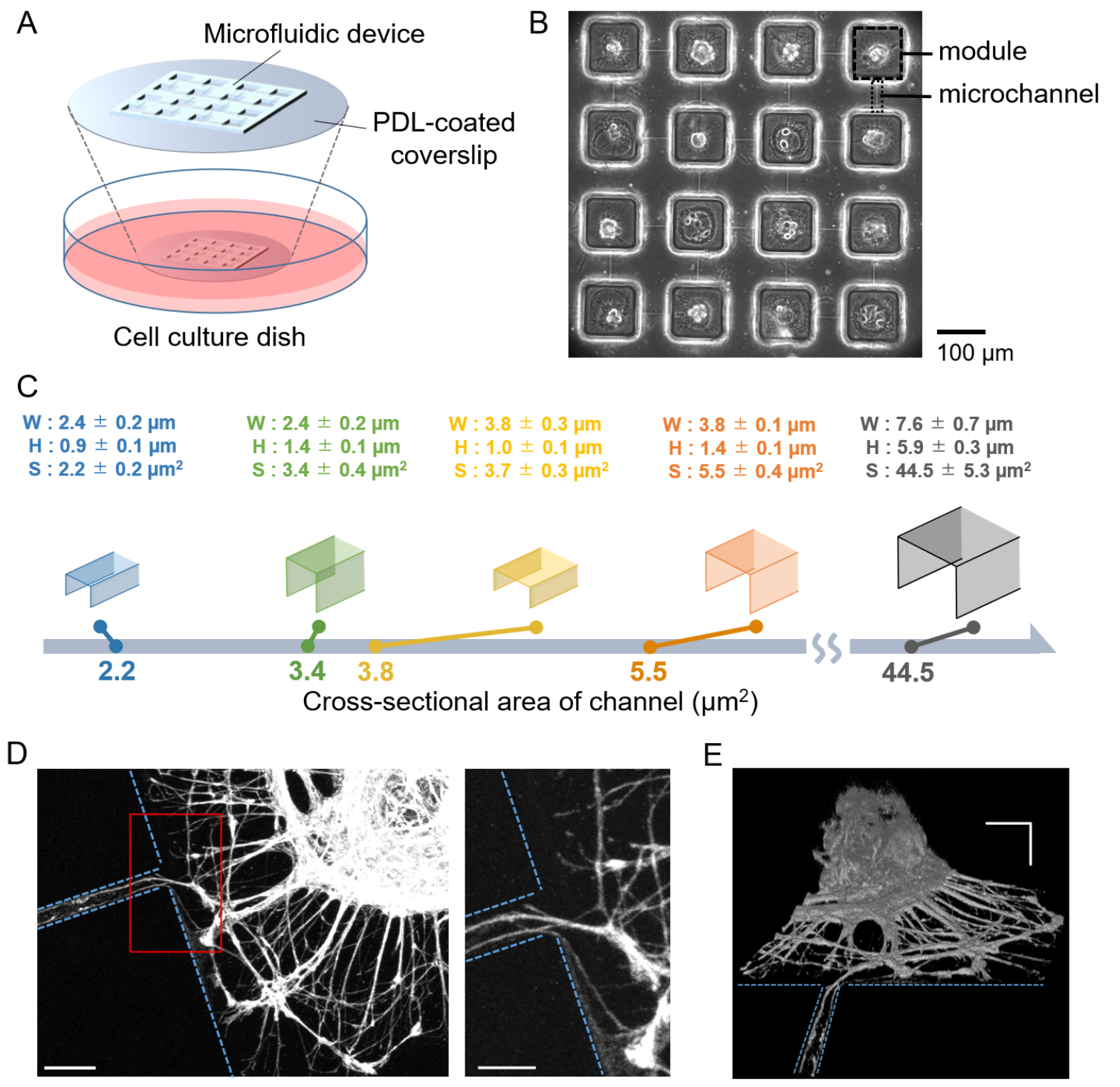

Figure 2. Neuronal culture. (A) A schematic of the engineered networks. The microfluidic device was attached to a PDL-coated coverslip, and rat cortical neurons were plated on it. (B) Phase-contrast micrograph of the engineered networks at 14 days in vitro (DIV). (C) A schematic of the microchannels and their geometry. W: width, H: height, S: cross-sectional area. Mean±SD. (D) Confocal fluorescence micrograph of the engineered network. Right: a magnified image of the area surrounded by the red line in the left panel. The blue dashed line represents the boundaries of microfluidic device. The scale bars on left and right panel are 10

and 5 μm, respectively. (E) 3D image reconstructed from cross-sectional micrographs. The horizontal and vertical scale bars are both 10 μm.

## 2.2. Suppression of excessive synchrony in small-channel networks

To assess and compare the dynamics of neuronal networks in small-channel and conventional-channel devices, we used fluorescence calcium imaging to record spontaneous neuronal activity and analyzed it in terms of synchrony and neuronal ensembles. Neurons were transfected with the fluorescent calcium probe GCaMP6s, and spontaneous activity was recorded for 30 min. The instantaneous spike rate of each neuron was inferred from the fluorescence intensity by using the deep learning-based algorithm CASCADE.[28]

A fluorescence micrograph, fluorescence traces of neurons, and an inferred spike rate of a neuron are shown in Figs. 3A and B for a small-channel network with a channel size of 3.7 μm$^2$. A majority of the neurons comprising the network were active; notably, the timing of their activation was diverse. First, to evaluate the degree of neuronal synchrony in the networks, we calculated correlation coefficients (CCs) between each neuron based on their spike rates (Fig. 3C). Only a small fraction (6.65%) of the neuron pairs had CCs greater than 0.8 ($n$ = 496 neuron pairs), and the mean CC decreased with channel size (Fig. 3D). The decrease in the mean CC was not caused by changes in the mean spike rate, as there were no significant differences observed for each channel size (Fig. S2).

Subsequently, we investigated how the physical separation between two neurons influences their correlation. For this analysis, we defined the *distance* between two neurons as the number of channels that connect their affiliated modules along the shortest path. We found that the CC decreased with increasing distance for all of the microchannel sizes and that the dependence was stronger for networks with smaller microchannels. However, it was interesting that the 3.7 μm$^2$ channel exhibited a stronger trend in the distance dependence of CC compared to the 3.4 μm$^2$ channel (Figs. 3E and S3). For neuron pairs in the same module (distance = 0), the CC was highly independent of the channel size. These results suggest that synchrony between a pair of neurons is constrained by the number of microchannels that couple the pair, and the distance-synchrony relationship can be mathematically modeled by considering a simple probabilistic model of activity propagation in microchannels (see Supplementary Material).

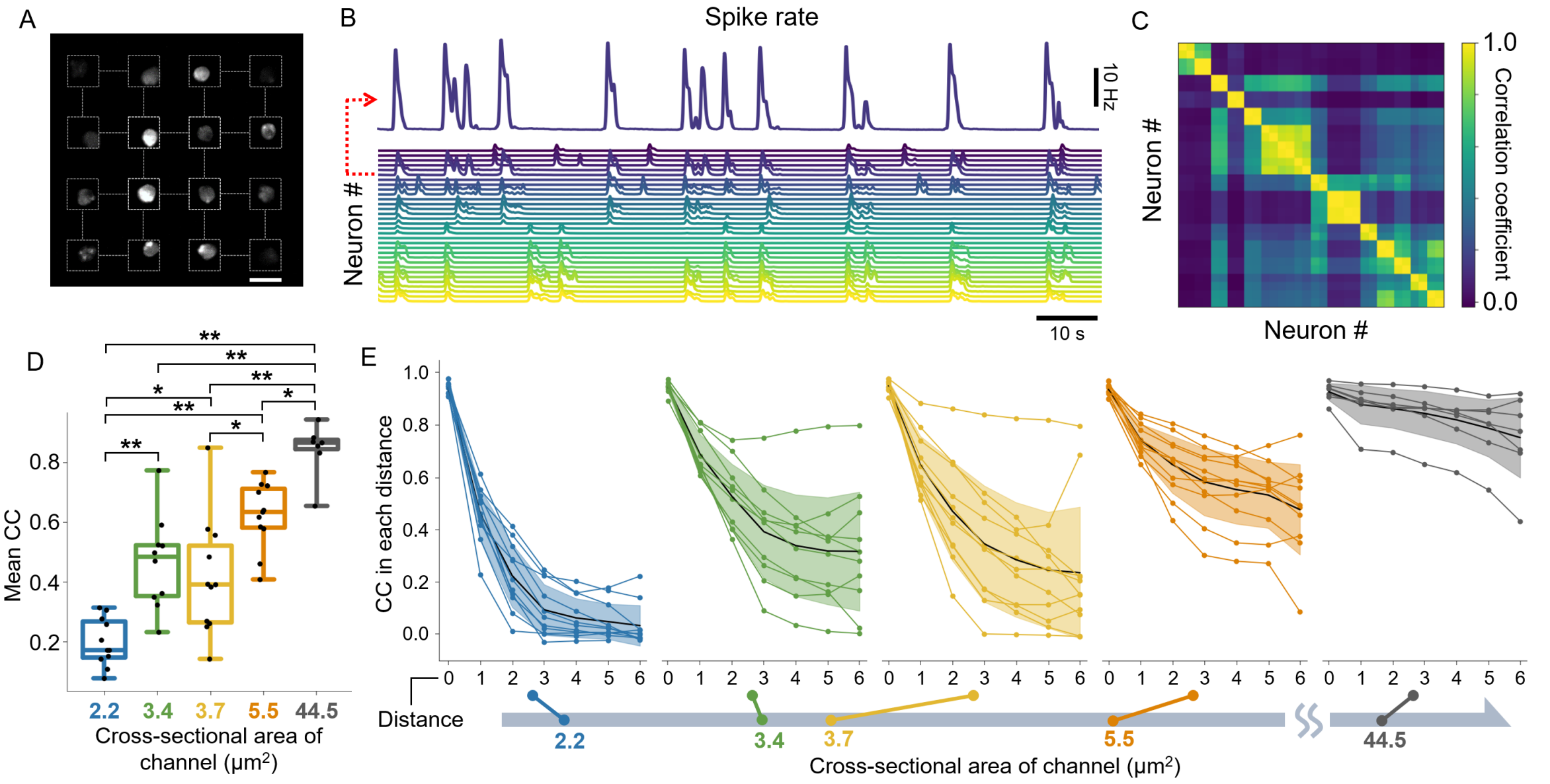

Figure 3. Suppression of excessive synchrony in small-channel networks. (A) A fluorescence micrograph shown in SD during calcium imaging. The dashed line indicates the contour of the modular network. The scale bar is 100 μm. (B) Estimated spike rate during the recording of spontaneous activity. Top: a representative neuron. Bottom: all neurons of a single network. (C) Pairwise correlation coefficients between neurons computed from the estimated spike rate. The depicted data are for the network identical to that shown in Fig. 3B. (D) Comparison of the mean correlation coefficient (mean CC) in each network for channels with sizes of 2.2 μm$^2$ ($n$ = 11), 3.4 μm$^2$ ($n$ = 10), 3.7 μm$^2$ ($n$ = 11), 5.5 μm$^2$ ($n$ = 11), and 44.5 μm$^2$ ($n$ = 7). The boxes and whiskers represent the first/third quartiles and the range, respectively. *$p$<0.05 and **$p$<0.01 (Steel–Dwass's multiple comparison test). (E) CC against distance in each network (colored lines). Mean values at each distance are plotted (black lines), with shading indicating the SD.

### 2.3. Neuronal ensembles in engineered networks

To assess the diversity of spatiotemporal patterns in spontaneous neural activity, we detected neuronal ensembles in engineered neuronal networks (Figs. 4A-D; see Methods section for details) and analyzed their properties. Fig. 4E shows a raster plot of a network grown in a small-channel device (3.7 μm$^2$), along with four detected neuronal ensembles. Statistically, one to six ensembles were detected in small-channel devices, with networks with smaller channels tending to exhibit a larger number of ensembles (Fig. 4F). In contrast, networks with the conventional channel (44.5 μm$^2$) exhibited a single ensemble in which all of the neurons were coactive (Fig. S5), except for one sample that exhibited two ensembles ($n$ = 7).

We further found that multiple ensembles detected in small-channel devices exhibited greater dissimilarity than those detected in conventional-channel devices. This finding was quantified by calculating the cosine similarity between all of the *population vectors* (see the Neuronal Ensemble Analysis section in Methods) associated with ensembles in each network, which showed that population vectors in different ensembles were clearly distinct (Fig. 4G).

A comparison of similarities between ensembles demonstrated that networks with 2.2 and 3.7 μm$^2$ channels displayed a more distinct combination of ensembles than those with conventional channels (Fig. 4G). Interestingly, the mean similarity of population vectors in the 3.4 μm$^2$-channel (2.4 μm in width, 1.4 μm in height) networks was not significantly lower than that in the conventional-channel networks, despite the cross-sectional area being smaller than that in the 3.7 μm$^2$ channels (3.8 μm in width, 1.0 μm in height) (Fig. 4H). The variation in neuronal ensembles exhibited by a network may depend more on the channel height rather than on its width in microfluidic devices. In summary, small-channel networks, in particular, the 2.2 and 3.7 μm$^2$-channel networks, display a greater diversity of neuronal ensembles in spontaneous activity.

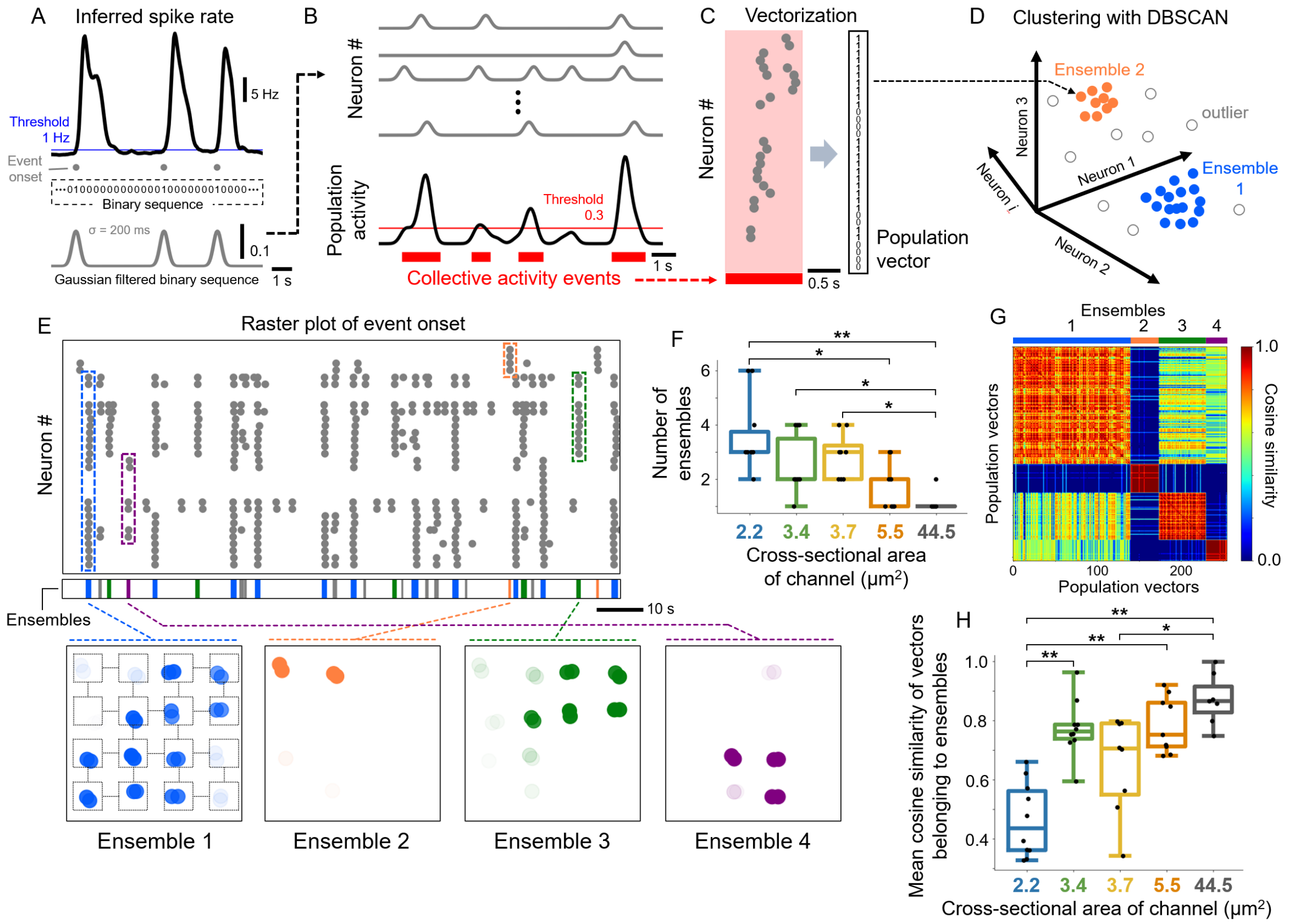

Figure 4. Neuronal ensembles in engineered networks. (A) Burst event detection from spike rate and filtering. Event onset was defined as the time point at which the spike rate exceeded the threshold (1 Hz). A binary sequence is created by setting it to 1 for frames with an event onset and to 0 for all of the other frames. The binary sequence was filtered with a Gaussian distribution (sigma = 200 ms) for subsequent collective activity detection. (B) Detection of collective activity events. Each filtered event was collected along with neurons, and a threshold (0.3) was applied to detect coactivity. (C) Vectorization of coactivity. Events during the coactivity time scale were binarized (event: 1, no event: 0) in each neuron. (D) Schematic illustration of clustering with DBSCAN for detecting neuronal ensembles. The population vectors have the dimension of the number of neurons. Each dot denotes a population vector.

All of the population vectors in each sample were classified by DBSCAN. Each cluster corresponds to a neuronal ensemble. Vectors not classified into a cluster were considered to be outliers. (E) Raster plot of event onset and detected neuronal ensembles in a representative network (channel size is 3.7 µm). Top: raster plot of events and detected neuronal ensembles corresponding to the raster plot. The color indicates different ensembles, and gray indicates outliers. Bottom: spatial map of neurons belonging to each ensemble. The gradation of each neuron color indicates the contribution rate to the ensemble. (F) Number of detected neuronal ensembles in each network. The sample sizes of each channel were 2.2 $\mu m^2$ ($n$ = 10), 3.4 $\mu m^2$ ($n$ = 10), 3.7 $\mu m^2$ ($n$ = 8), 5.5 $\mu m^2$ ($n$ = 9), and 44.5 $\mu m^2$ ($n$ = 7). *$p<0.05$ and **$p<0.01$ (Steel–Dwass's multiple comparison test). (G) Cosine similarity matrix of population vectors belonging to neuronal ensembles in the representative network. The colored line above the similarity matrix indicates the range of population vectors corresponding to each ensemble. (H) Comparison of the mean cosine similarity of population vectors belonging to ensembles in each network. Each dot denotes the mean cosine similarity from one sample. *$p<0.05$ and **$p<0.01$ (Steel–Dwass's multiple comparison test).

### 2.4. Induced activities differed according to the location of the stimulated neuron.

To investigate the mechanisms underlying the emergence of diverse neuronal ensembles in small-channel networks, we used optogenetic tools to stimulate neurons in individual modules and examined the spatial patterns of evoked activity. The recording of the stimulation-evoked activity was conducted at 14-15 DIV. To deliver optogenetic stimulation, neurons were virally transfected with the redshifted channelrhodopsin ChrimonR[29] at 4 DIV, and the neurons were irradiated with red LED light via patterned illumination by using a digital mirror device.[30,31] The illumination area was defined as the rectangle surrounding the neurons in each module (Fig. 5B). Following a previous study,[32] the patterned illumination was irradiated for a duration of 4 s at a duty ratio and repetition rate of 5% and 10 Hz, respectively. Sixteen modules were stimulated in the order illustrated in Fig. 5A, and an interval of 10 s was set before stimulating the next module. The whole sequence of stimulating all 16 modules was repeated eight times.

The obtained stimulus responses were analyzed based on bursting events detected from the CASCADE-inferred spike rates. As shown in Fig. 5C, the spike rate of the targeted neurons increased at the onset of stimulation, and the neurons remained active until the end of stimulation. The event onset of burst event was detected following the algorithm described in the previous section. The event was considered to be evoked by the stimulation if the event onset was initiated between 0.5 s before the start of photoirradiation and 0.5 s before the end of photoirradiation, as the inferred spike rate is a Gaussian-convolved signal. The spatial pattern of the evoked activity was analyzed by vectorizing the detected onsets following the method described for the neuronal ensemble analysis (Fig. 5D).

A raster plot of a small-channel network (2.2 μm$^2$) showed that stimulation of the different modules induced different activities (Fig. 5E). To spatially map the spatial extent of the evoked activity, corresponding to neuronal ensembles, we calculated the probability for each neuron to become active upon the stimulation of one module (Fig. 5F, G). The response probability upon stimulation of module 5 is shown in Figure 5G, which shows that only the neurons in the stimulated module and its neighbors are activated. The response probabilities for all of the modules are summarized in Figure 5H, which clearly shows that the induced activities were localized and that the stimulation of different modules evoked different activity patterns. Comparison of the mean cosine similarity of population vectors derived from stimulating each module demonstrated that, although the repetitive stimulation of identical modules evoked similar activity patterns, activities induced by the stimulation of different modules were dissimilar (Figs. 5I and J; $n = 4$).

Interestingly, stimulating certain modules was found to elicit similar stimulus-evoked activities, such as the group surrounded by a dashed magenta line in Figs. 5H and I. The neurons in this group are expected to be more strongly coupled than the others. The number of such module groups may reflect the diversity of neuronal ensembles that emerges in spontaneous activity. In summary, neuronal ensembles generated by neurons in identical modules are stable, and diverse neuronal ensembles observed in the spontaneous activity of small-channel networks are likely to emerge due to the intrinsic initiation of activity from different locations.

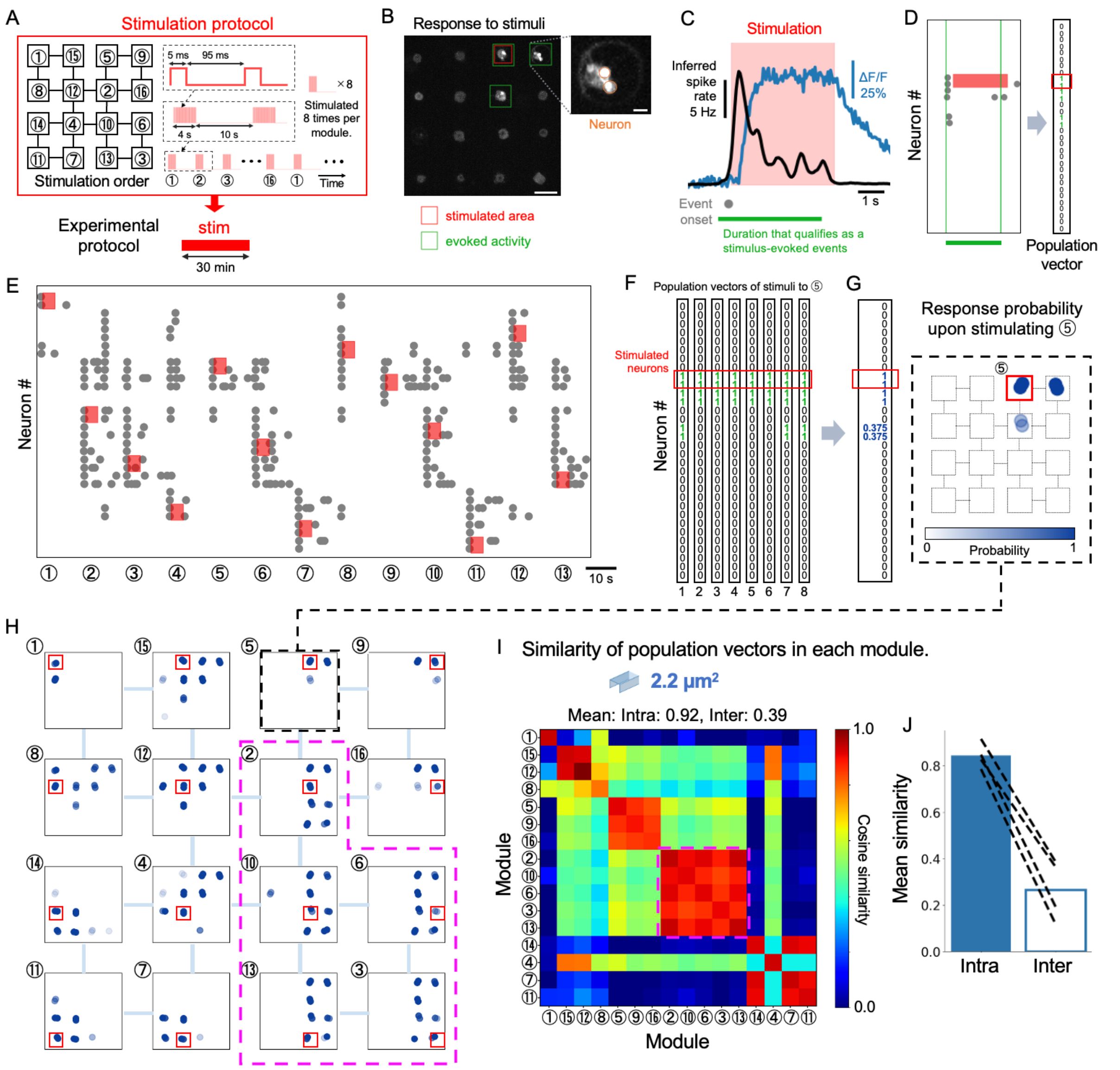

Figure 5. Stimulation of different modules induced various responses in small-channel networks. (A) Experimental protocol. The experiment consisted of only a 30 min stimulation session (bottom). During the session, each module was stimulated according to the stimulation order (top left). The stimulated area was defined as a rectangle to cover all of the neurons in the module. For each stimulation, a 5 ms light pulse with 10 Hz was applied for 4 s. The duration between each stimulation and a module was 10 s. (B) A fluorescence micrograph shown in SD for a stimulation response (scale bar, 100 µm). The area surrounded by the red rectangle is the stimulated module. The areas surrounded by bright green responded to the stimulation. A magnified image of a single module is shown in the inset, with neurons expressing GCaMP6s marked with orange circles (scale bar, 20 µm). (C) Neuronal response to stimulation (blue line: relative fluorescence of the GCaMP6s signal; black line: estimated spike rate; gray dot: event onset detected from the spike rate). We defined the duration that qualifies as a stimulus-induced event for the following analysis (bright green line). (D) Stimulus-induced events were transformed into the population vector. (E) Raster plot of events in a representative network. The gray dots indicate the events. The red rectangles indicate the neurons and the duration of stimulation. The numbers below the raster indicate the order of the stimulated module. (F) A collection of population vectors in response to the stimulation of a single module. The area surrounded by the red rectangle indicates the responses of the stimulated neurons. (G) Response probabilities were calculated via the sum

of events divided by 8 in each neuron (left) and displayed as a spatial map (right). The red rectangle indicates the stimulated module. (H) Spatial maps of the response probabilities of all of the modules. The maps surrounded by the magenta dashed line correspond to the similarity map (I). (I) Similarity of population vectors sorted and averaged by module. Each cell in the matrix indicates the mean similarity of population vectors in a pair of modules. (J) Mean similarity of intramodules and intermodules in each network (dashed black lines). Bar height indicates the mean value.

**2.5. Stimulus-induced plasticity in engineered neuronal networks**

Accumulating evidence suggests that the plasticity of neuronal ensembles is related to learning and memory processes, which are among the most fundamental functions of the brain.[5,33] To evaluate this phenomenon in cultured neuronal networks, we investigated whether spontaneous neuronal ensembles change significantly in response to repetitive synchronous stimulation of multiple neurons. The same photostimulation protocol as that used in Section 2.5 was used for this experiment (see Fig. 5A) and was repetitively applied for 30 min to neurons of four center modules of the small-channel networks (Fig. 6A). Spontaneous activity was recorded before and after the synchronous stimulation, and the statistics were compared. Based on the properties of the neuronal ensembles that were analyzed in Section 2.4, we analyzed the data according to two groups, which included the shallow-channel (2.2 and 3.7 $\mu m^2$) and tall-channel (3.4 and 5.5 $\mu m^2$) networks.

Synchronous photostimulation evoked activity not only in the targeted neurons but also in some off-target neurons (Fig. 6B). Moreover, the response probabilities tended to decrease with increasing distance from the photoirradiation. An example of evoked activities in a representative network is presented in Fig. 6C. To quantify the spatial extent of activity evoked by synchronous stimulation, we computed the response probability for each neuron. Analysis revealed that the response probability decreased in the order of targeted neurons, neighboring neurons, and nonneighboring neurons (Fig. 6D). This result is consistent with observations in the stimulation of a single module (Section 2.4), wherein the spread of activity was spatially limited. The nonstimulated modules in shallow-channel networks, whether neighboring or nonneighboring, tended to exhibit a lower response probability than did the tall-channel networks. Therefore, even when the stimulation area was expanded to include four modules, the activity propagation remained spatially uneven.

A comparison of spontaneous activity recorded before and after synchronous stimulation demonstrated that stimulation indeed changed the neuronal ensembles embedded in the network (Figs. 6E-H). During synchronous stimulation, an ensemble that spans the entire network, such as Ensemble 1 in the example shown in Fig. 6E, tended to dominate other localized ensembles. A comparison of the spontaneous activity recorded before and after stimulation demonstrated that both the number (Fig. 6F) and proportion (Fig. 6G) of neuronal ensembles changed. In contrast, when stimulation was absent, the ensembles remained roughly stable throughout the experiment (Fig. S6).

Finally, we quantified the degree of change in ensembles by evaluating the mean cosine similarity of the population vectors that were observed in the spontaneous activity (Fig. 6H). The mean value of the "pre" and "post" submatrices represents the similarity among the observed population vectors. Taking their ratio against the mean value in the "pre vs. post" submatrix, the relative change in population vectors is obtained, with a value of one signifying no change in the population vector and greater than one signifying a change.[34] The results that were obtained for networks that received repetitive synchronous stimulation during the stimulation phase (shallow channel: $n = 14$, tall channel: $n = 15$) and those that did not (shallow channel: $n = 11$, tall channel: $n = 14$) are summarized in Fig. 6I. Importantly, in networks with shallow channels, the relative change in the stimulated networks was significantly greater than that in the nonstimulated networks. Moreover, such a trend was observed only in the shallow-channel networks and not in the tall-channel networks (Fig. 6I). These results show that neuronal ensembles in engineered networks, especially in networks with shallow channels, could be manipulated by external stimulation, suggesting that they represent a novel in vitro model system for studying the properties of neuronal ensembles.

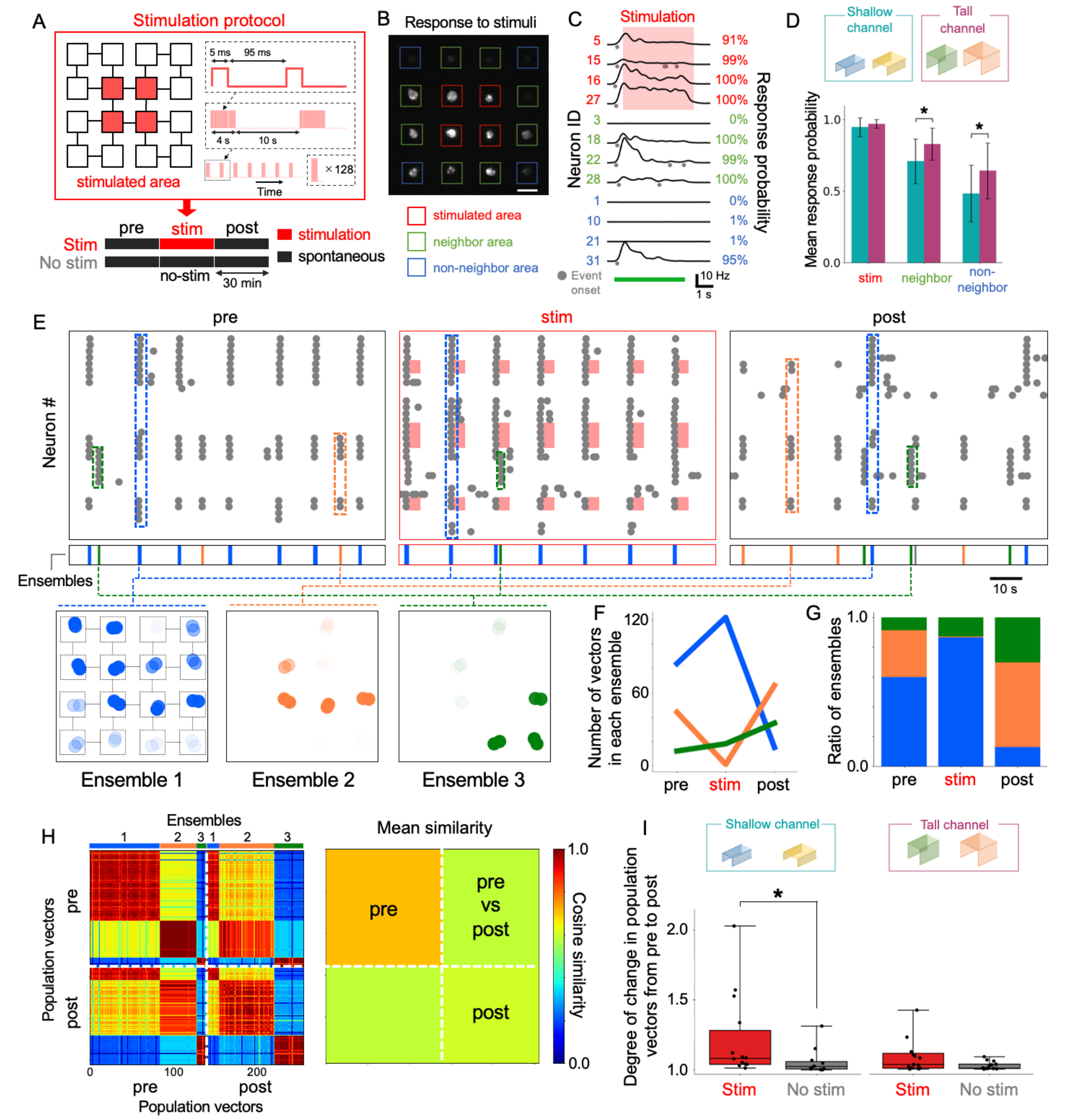

Figure 6. Stimulus-induced plasticity of neuronal ensembles in engineered networks. (A) Experimental protocol. Bottom: session design. There were two experiments: a stimulation experiment (Stim) and a control experiment (No stim). Both experiments consisted of three sessions, and the first and third sessions involved recording spontaneous activity. In the Stim experiment, the second session consisted of stimulating the networks and recording the activity. In the no-stim experiment, the second session involved recording spontaneous activity. All of the sessions lasted 30 min. Top: stimulation protocol that was used in the stimulation session. Four modules located in the center were stimulated areas. All of the neurons belonging to these modules were stimulated simultaneously. The applied stimulation protocol was the same as that described in Section 2.4, and total stimulation was 128 times in 30 min. (B) A fluorescence micrograph shown in SD for a stimulation response. The modules were categorized into three groups based on their locations: the stimulated areas (red squares), the areas neighboring the stimulated areas (green squares), and regions that did not belong to either category (blue squares). The scale bar represents 100 µm. (C) Representative neuronal responses to stimulation. Each trace (black line) shows the spike rate of a neuron belonging to the stimulated, neighboring, and nonneighboring areas from top to bottom during stimulation.

The duration of stimulation and the stimulated neurons are indicated by the light red rectangle. The number and percentage displayed on the left side of the spike rate indicate the neuron ID and the probability of being evoked by stimulation, respectively. Response probabilities were calculated by whether events (gray dot under the spike rate) occurred during the duration (displayed in bright green below) of the stimulation session. (D) Comparison of the mean response probability in the shallower-channel ($n$ = 14) and taller-channel ($n$ = 15) networks. The mean response probability was calculated for each group of locations (stimulated, neighboring and nonneighboring areas) of a network. The height of the bar and the error bar indicate the mean and SD of the mean response probability, respectively. *$p$<0.05 (two-tailed Student's t-test). (E) Raster plot of event onset and detected neuronal ensembles in a representative network. Top left: prestimulation session. Top middle: stimulation session. The light red rectangles indicate the neurons and the duration of stimulation. Top right: poststimulation session. Bottom: Spatial map of neurons belonging to neuronal ensembles. The gradation of a neuron indicates the contribution ratio to the ensemble. (F) Number of population vectors belonging to ensembles in each session of the network. (G) Proportion of population vectors belonging to ensembles in each session of the network. (H) Left: the cosine similarity matrix of population vectors belonging to ensembles during pre- and poststimulation sessions of the network. The bar above the matrix indicates vectors belonging to each ensemble. Right: mean cosine similarity calculated for each group (pre or post) and between groups. (I) Degree of change in population vectors from pre to post. The degree was evaluated as the ratio of the mean intragroup similarity against the mean intergroup similarity. Left: the results for shallow-channel networks (Stim: $n$ = 14, No stim: $n$ = 11). Right: the result of tall-channel networks (Stim: $n$ = 15, No stim: $n$ = 14). The boxes and whiskers represent the first/third quartiles and the range, respectively. Each black dot represents one network. ns: not significant, *$p$<0.05 (Mann–Whitney U test).

## 3. Discussion

In the current study, we used microfluidic devices to construct a biological neuronal network with a hierarchically modular structure as an experimental prototype for analyzing neuronal ensembles in vitro. The cell engineering technology presented here can be scaled to flexibly design other network structures by varying the number of modules and combining multiple types of microchannels and connection paths. Alternative approaches to suppressing synchronous activity in cultured networks include the Modular Neuronal Network (MoNNet) method, which exploits neuronal self-organization to construct modular neuronal networks.[14] The MoNNet has proven to be useful for modeling brain disorders. Although the present microfluidics-based approach adds some complexity to device fabrication, it allows for controlled manipulation of network connectivity, thus enabling us to model various nonrandom features in brain networks.[18]

Our experiments demonstrated that the neural correlation and diversity of neuronal ensembles in cultured cortical networks could be controlled by precisely adjusting the microchannel geometry. Naturally, neural correlation depends on the number of connections between neuronal populations both in biological neurons and in mathematical models.[22, 11, 36] One nontrivial observation in the current study was that the diversity of neuronal ensembles in engineered networks was not solely determined by the cross-sectional area of the microchannels. We found that networks with shallow-and-wide microchannels exhibited more diverse neuronal ensembles than those with tall-and-narrow microchannels. We speculate that such difference was caused by a lower number of neurites in the shallower microchannels, even when they possessed an equivalent cross-sectional area. Peyrin et al. showed that the number of axons within a microchannel remained relatively stable (between approximately 2 and 5 axons) even when the width was reduced from 4 to 2 μm at the same height (3 μm).[15] Conversely, Ristola et al. demonstrated a significant decrease in the number of axons (from approximately 20 to 5 axons) when the height of the microchannel was reduced from 3.5 to 1 μm for the same width (10 μm).[17] Based on these findings, it is plausible that the shallower microchannels contained fewer neurites and that these networks exhibited more diverse neuronal ensembles due to the reduced connectivity between neuronal populations.

In the shallow-channel networks, repetitive optogenetic stimulation induced plastic changes in neuronal ensembles (Fig. 6). The formation of new ensembles by repetitive stimulation of specific neurons has been demonstrated both in vivo and ex vivo,[32,37] which aligns with our observations. Synaptic plasticity, such as Hebbian plasticity and spike-timing-dependent plasticity, and the intrinsic plasticity of neurons have been regarded as being underlying mechanisms behind the formation of neuronal ensembles.[38,39,32] It is plausible that weak intermodular coupling in shallow-channel networks may have facilitated changes in coupling strength through various forms of plasticity, resulting in alterations in neuronal ensembles. Future challenges include the identification of the underlying rules driving these changes in our engineered networks, as well as the precise imprinting of specific ensembles. The identification of such mechanisms would not only contribute to understanding memory and learning in biological systems but also lead to a fundamental technology for realizing adaptive physical reservoir computing based on biological neural networks in vitro.[30,40]

## 4. Conclusion

In this study, we fabricated biological neuronal networks with controlled neuronal connectivity by using microfluidic devices. The results showed that networks with sparse intermodular connections exhibit diverse neuronal ensembles in spontaneous activity and that the statistics of these ensembles significantly change after repeated external stimulation of the network. Networks possessing shallower microchannels exhibited increased ensemble diversity, and their neuronal ensembles were more sensitive to stimulation. Our study demonstrated that precision neuroengineering technologies based on microfluidic devices can be coupled with fluorescence calcium imaging and optogenetics tools to provide a versatile framework for studying the mechanisms of plasticity and stability of neuronal ensembles under well-defined conditions. This technology may also contribute to modeling neurological disorders in vitro and realizing physical reservoir computing based on biological neurons. Future research should focus on identifying the underlying mechanisms of the plasticity of neuronal ensembles and their requirements, which will enhance the understanding of the learning and memory processes of biological neuronal networks.

## 5. Methods

*Master mold fabrication*: Microfluidic devices were fabricated following a previously published protocol,[20] with some modifications. First, the master mold for structuring the PDMS was fabricated by patterning two layers of SU-8 photoresist. To form the first layer, which contained the patterns for both reservoirs (modules) and microchannels, a mixed solution of SU-8 3005 (Kayaku Advanced Materials) and γ-butyrolactone (GBL; FUJIFILM Wako Pure Chemical) was first spin-coated onto a 2-inch silicon wafer at 3000 rpm and then baked for 1 and 5 min at 65 °C and 95 °C, respectively. The ratio of SU-8 to GBL was varied, as noted in the Results section, to achieve different thicknesses of the first layer. Afterward, the sample was exposed to UV light through a photomask by using a mask aligner (SUSS MJB4) and subsequently baked for 1 and 3 min at 65 °C and 95 °C, respectively. The pattern was developed in an SU-8 Developer and rinsed in 2-propanol. To fabricate the second layer, SU-8 3050 was spin-coated onto the first layer at 1500 rpm and was baked for 1 and 30 min at 65 °C and 95 °C, respectively. Photolithography was then performed by using a photomask with only the reservoir pattern, followed by a postexposure bake for 1 and 5 min at 65 °C and 95 °C, respectively; finally, the pattern was developed as in the first layer.

The microfluidic device was then fabricated via replica molding by using the master mold. The base and curing agent of Sylgard 184 (PDMS) were mixed at a ratio of 7.5:1 and degassed by using a vacuum chamber. The mixture was drop-cast on the mold and baked for 2 h at 70 °C. After baking, the cured PDMS was peeled off from the mold, subsequently cleaned by sonication in 100% EtOH and deionized water, and then placed under UV light for more than 30 min for sterilization.

*Master mold characterization*: The mold was first coated with a thin layer of Pt (< 20 nm) using an ion-coater. The 3D topography of the master mold was then measured by using confocal microscopy (Keyence VK-X260), and the data were analyzed by using MultiFileAnalyzer software (Keyence).

*Cell culture*: All of the procedures involving animal experiments and gene transfection experiments were approved by the Tohoku University Center for Laboratory Animal Research (2020AmA-001) and Tohoku University Center for Gene Research (2019AmLMO-001). Primary rat cortical neurons were obtained from embryonic day 18 rats according to a previously published protocol.[41] Before cell culture, the microfluidic device was attached to a PDL-coated coverslip filled with neuronal plating medium [MEM (Gibco 11095-080) + 5% fetal bovine serum + 0.6% D-glucose] and placed in a vacuum chamber to remove air bubbles entrapped in the microfluidic device. The medium was then completely replaced with fresh plating medium, and the coverslip was stored in a $CO_2$ incubator (37 °C, 5% $CO_2$) for at least 3 h. Rat cortical neurons were then plated on the coverslip and cocultured with astrocyte feeder cells in 5 mL of N2 medium [MEM (Gibco 11095-080) + 10% N2 supplement + 1% ovalbumin (Sigma 17504-044) + 1 M HEPES]. At 4 DIV, half of the medium was removed, and the cultured neurons were transfected with adeno-associated viral vectors encoding the fluorescent calcium probe GCaMP6s (Addgene 100843-AAV1) and the redshifted channelrhodopsin ChrimsonR (Addgene 59171-AAV1) by adding 3 and 2 μL, respectively, of the AAV solution to 2.5 mL of medium. At 5 DIV, the medium that was removed at 4 DIV was added, together with 1 mL of Neurobasal Plus medium [Neurobasal Plus Medium (A3582901, Gibco) + 2% B-27 Plus Supplement (A3582801, Gibco) + 0.25% GlutaMAX-I (35050-061, Gibco)]. Half of the medium was replaced with Neurobasal Plus medium at 8 DIV.

*Confocal imaging*: For confocal imaging of neurites penetrating the microchannels, NeuO dye (ST-01801, STEMCELL Technology) was added to the culture medium at a concentration of 0.2 μM, and the cells were incubated for 30 minutes. After rinsing with DPBS, the coverslip was immersed in a mixed solution containing 4% PFA and 4% sucrose for 15 minutes and then rinsed twice with DPBS. Finally, the coverslip was mounted on a glass slide by using ProLong Glass (P36982 Invitrogen) and stored in the dark at room temperature for 18-20 h.

Confocal imaging of the neurons was performed by using an inverted microscope (Olympus IX83) equipped with the MAICO MEMS confocal unit (C15890 Hamamatsu Photonics; laser wavelength: 488 nm). As the neurons expressed GCaMP6s, with excitation and fluorescence wavelengths being close to those of NeuO, the obtained images contain fluorescence from both NeuO and GCaMP6s. The images were analyzed by using HCImage software (Hamamatsu Photonics) and ImageJ software (NIH).

*Calcium imaging*: Neuronal activity was recorded at DIV 14 or 15 via fluorescent calcium imaging. The coverslip containing cultured neurons was rinsed with HEPES-buffered saline (HBS) (128 mM NaCl, 4 mM KCl, 1 mM $CaCl_2$, 1 mM $MgCl_2$, 10 mM D-glucose, 10 mM HEPES, and 45 mM sucrose) and transferred to a glass-bottom dish (3960-03535, Iwaki) filled with fresh HBS. The fluorescence intensity of the calcium indicator GCaMP6s was imaged by using an inverted microscope (IX83, Olympus) equipped with a 20× or 10× objective lens (numerical aperture, 0.80 or 0.40, respectively), an LED light source (Lambda HPX, Sutter Instrument), a scientific complementary metal-oxide semiconductor camera (Zyla 4.2P, Andor), and an incubation chamber (Tokai Hit). All of the recordings were performed at 37 °C and 20 frames/s by using Solis software (Andor).

*Optogenetic stimulation*: The neuronal networks were stimulated by activating ChrimsonR with patterned red light. Red light was delivered from a red LED (Solis 623-C Thorlabs) and through a digital micromirror device (DMD; Mightex Polygon400G) to stimulate specific areas in the network. Stimulation patterns were generated by using PolyScan2 software (Mightex).

*Image processing*: Image processing was performed with ImageJ software and custom-written Perl and Python codes. To extract neuronal activity, regions of interests (ROIs) representing neurons (2 neurons per module) were manually selected from a network, and the mean intensity within the ROIs was calculated. The relative fluorescence (RFU, $\Delta F/F$) of each cell was calculated as $(F\text{-}F_0)/F_0$, where $F$ and $F_0$ denote the mean and baseline intensities, respectively. The RFU was converted to the spike rate by using the CASCADE algorithm, which is a deep learning-based method for spike inference from calcium dynamics.[28] The deep learning model was trained by using a dataset from the literature, which contained simultaneous patch-clamp and calcium imaging data. The calcium imaging signal was adjusted to align with the frame rate and noise level of the RFU (= 1) that were similar to the conditions in our experiments.

*Correlation analysis*: The functional correlation between neurons was calculated by using the Pearson correlation coefficient, as described below.

$$r_{ij} = \frac{\sum_t [x_i(t) - \overline{x_i}]\,[x_j(t) - \overline{x_j}]}{\sqrt{\sum_t [x_i(t) - \overline{x_i}]^2}\sqrt{\sum_t [x_j(t) - \overline{x_j}]^2}}, \quad (1)$$

where $x_i$ and $\overline{x_i}$ are the instantaneous and mean spike rates of neuron *i*, respectively. The mean of $r_{ij}$ across all of the neuron pairs was calculated by excluding the diagonal elements of the $r_{ij}$ matrix. Neurons with spike rates below 1 Hz were regarded as being nonactive and removed from the correlation analysis.

*Burst event detection*: Burst events in each neuron were detected by applying a threshold (1 Hz) to the spike rate. The time when the spike rate increased above the threshold was defined as the event onset. Events were counted as separate if the spike rate decreased below the threshold before rising again for the next event.

*Neuronal ensemble analysis*: Network activity patterns were classified into neuronal ensembles as follows:

(1) First, bursting events of each neuron were detected following the method described above, and a binary sequence of activity onsets was created for each neuron wherein 1 was set to the starting time of the events (Fig. 4A, top).

(2) Subsequently, the binary sequence was filtered with a Gaussian function (SD = 200 ms) (Fig. 4A, bottom), and the filtered sequence was summed across all of the neurons to obtain the population activity trace (Fig. 4B). Afterward, a threshold of 0.3, which corresponded to coactivation of more than four neurons, was applied to detect a collective activity event. The onset and offset of collective activity events were set to 0.2 s before and after the threshold crossing, respectively, to detect all of the neurons participating in the population activity.

(3) Each collective activity was subsequently transformed into a *population vector*, in which the element was set to 1 if the corresponding neuron was active during the period and set to 0 otherwise (Fig. 4C).

(4) Finally, neuronal ensembles, or repeatedly occurring activity patterns, were detected by using an unsupervised learning algorithm known as density-based spatial clustering of applications with noise (DBSCAN) (Fig. 4D). The algorithm clusters data points (population vectors) according to density. Unlike the *k*-means algorithm, the number of clusters is not specified as a classification parameter.[42] The cosine distance was used to calculate the distance between data points, and the distance threshold (epsilon) was set to 0.13. The minimum sample size was set to 5% of the number of data points in each network. Only the data points during the pre and post sessions were used for the analysis in Section 2.5. Therefore, the number of clusters consequently obtained corresponds to the number of ensembles in the network. The outlier points were considered noise and eliminated from the analysis of neuronal ensembles. The effect of epsilon on clustering performance is provided in Fig. S4, which shows that the conclusions derived in Section 2.3 are consistent and independent of the parameter.

*Similarity analysis*: The similarity between two population vectors $\boldsymbol{x} = [x_i]$ and $\boldsymbol{y} = [y_i]$ was defined as:

$$S_{xy} = \frac{\boldsymbol{x} \cdot \boldsymbol{y}}{|\boldsymbol{x}||\boldsymbol{y}|} = \frac{\sum_{i=1}^{n} x_i\, y_i}{\sqrt{\sum_{i=1}^{n} {x_i}^2}\sqrt{\sum_{i=1}^{n} {y_i}^2}}, \tag{2}$$

where $n$ is the number of neurons. For the evaluation of mean cosine similarity presented in Sections 2.3 and 2.5, the outlier population vectors were excluded from the analysis, and the mean was calculated over all pairs of population vectors that belonged to neuronal ensembles (Sections 2.3 and 2.5). In Section 2.4, mean similarity was calculated from all of the population vectors that were evoked in response to the stimulation of the modules.

## 6. Acknowledgements

We thank Mr. Iori Morita at the Fundamental Technology Center, Research Institute of Electrical Communication (RIEC), Tohoku University for the fabrication of the photomasks. The work was partly supported by MEXT Grant-in-Aid for Transformative Research Areas (A) and (B) "Multicellular Neurobiocomputing" (21H05164, 24H02332, 24H02334), JSPS KAKENHI (22H03657, 22K19821, 22KK0177, 23H00251, 23H02805, 23H03489), JST-CREST (JPMJCR19K3), the WISE Program for AI Electronics by Tohoku University, and the Cooperative Research Project Program of the RIEC, Tohoku University. This research was partly carried out at the Laboratory for Nanoelectronics and Spintronics, RIEC, Tohoku University.

Supporting Information

# Precision microfluidic control of neuronal ensembles in cultured cortical networks

*Hakuba Murota, Hideaki Yamamoto*, Nobuaki Monma, Shigeo Sato, and Ayumi Hirano-Iwata*

**Supporting Text**

*Mathematical modeling of the intermodular propagation probability.*

The relationship between *distance* and neuronal correlation can be described in a simplified mathematical model that considers the probability of activity propagation between modules (neuronal clusters) separated by microchannels. To do so, we define propagation probability $p$ as the probability that the activity of a neuronal cluster propagates to a neighboring cluster. Due to the fact that $p$ is independent for all of the microchannels, the probability for an activity to propagate to a module with a distance of $n$ will be $p^n$. Finally, a correction factor $a$ was multiplied to compensate for the variation in the probability at $n = 0$, and the dependence of the correlation coefficients on $n$ was fit against $ap^n$.

We found that this simple model fits the experimental results with high accuracy in all of the networks (coefficient of determination, $R^2 > 0.92$). This suggests that the propagation probability model can explain the behavior of activity propagation between modules. From the fitting curves, we estimated that $p$ depends on the cross-sectional area of the microchannels (Fig. S3D). Interestingly, $p$ tended to increase as the cross-sectional area increased. This result can be intuitively understood by considering that the number of axons coupling the modules increases in larger microchannels; thus, the probability for an activity to propagate increases.

## Supporting Figures

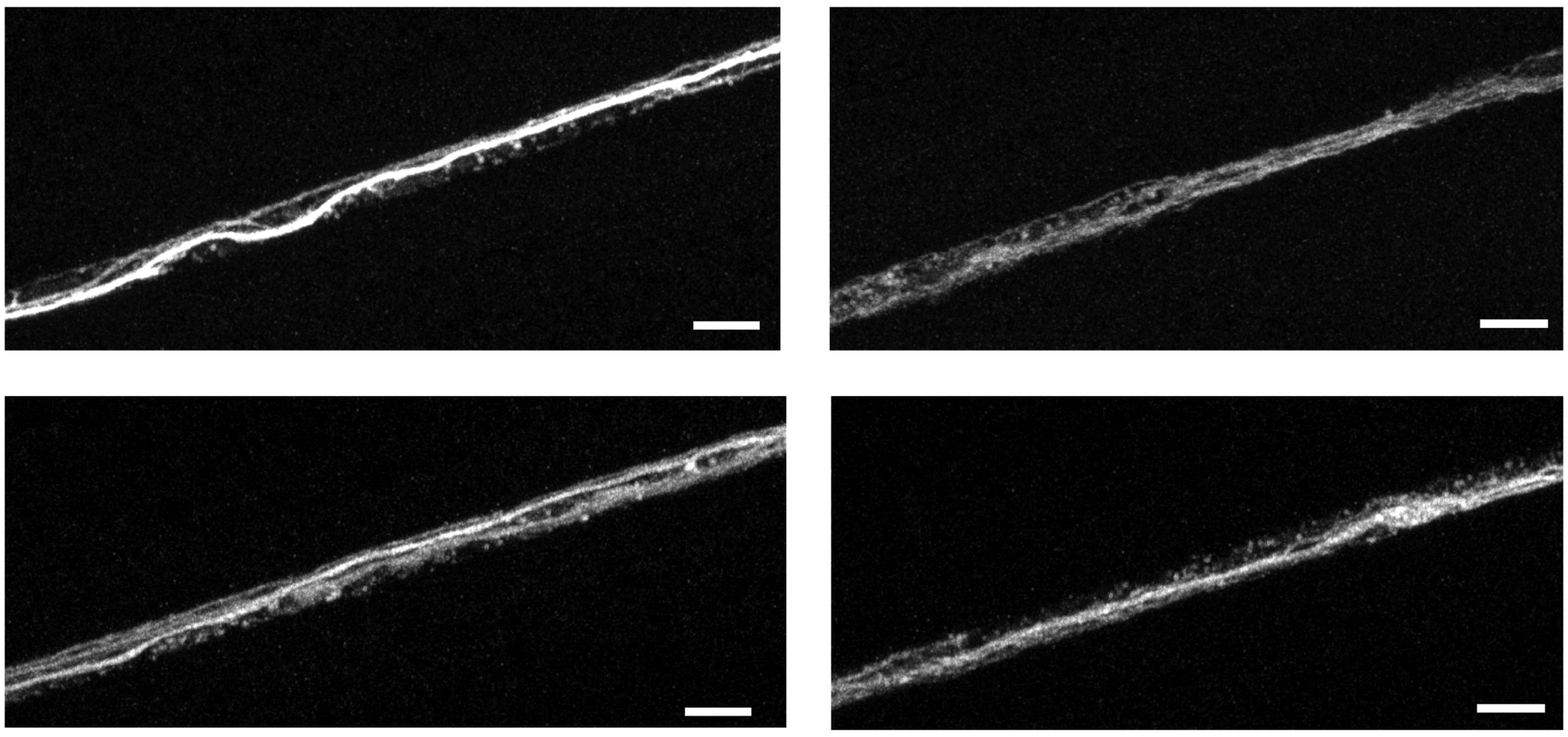

**Figure S1.** The neurites grew within the microchannel (2.2 μm$^2$). Four images are from different microchannels. The scale bars are 5 μm.

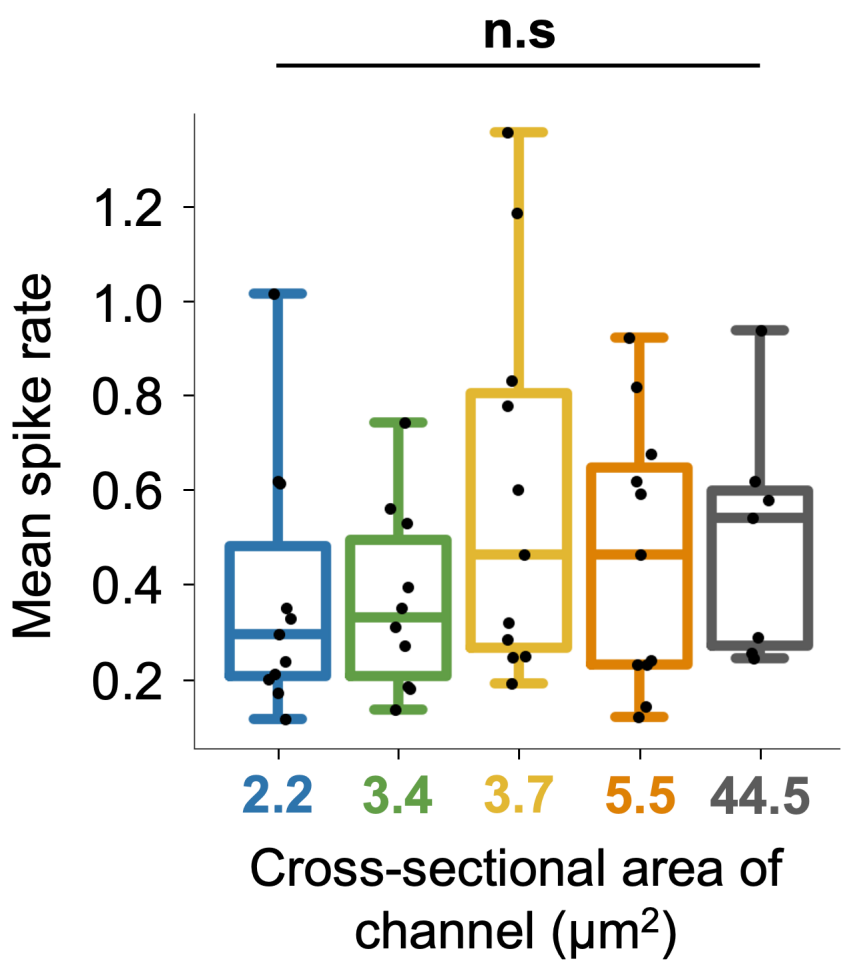

**Figure S2.** Mean spike rate in each network (black dots) summarized based on microchannel size. The mean spike rate was calculated by averaging the mean spike rate of each neuron in each network for channels with sizes of 2.2 μm$^2$ ($n$ = 11), 3.4 μm$^2$ ($n$ = 10), 3.7 μm$^2$ ($n$ = 11), 5.5 μm$^2$ ($n$ = 11), and 44.5 μm$^2$ ($n$ = 7). The boxes and whiskers represent the first/third quartiles and the range, respectively. n.s.: not significant (Steel–Dwass's multiple comparison test).

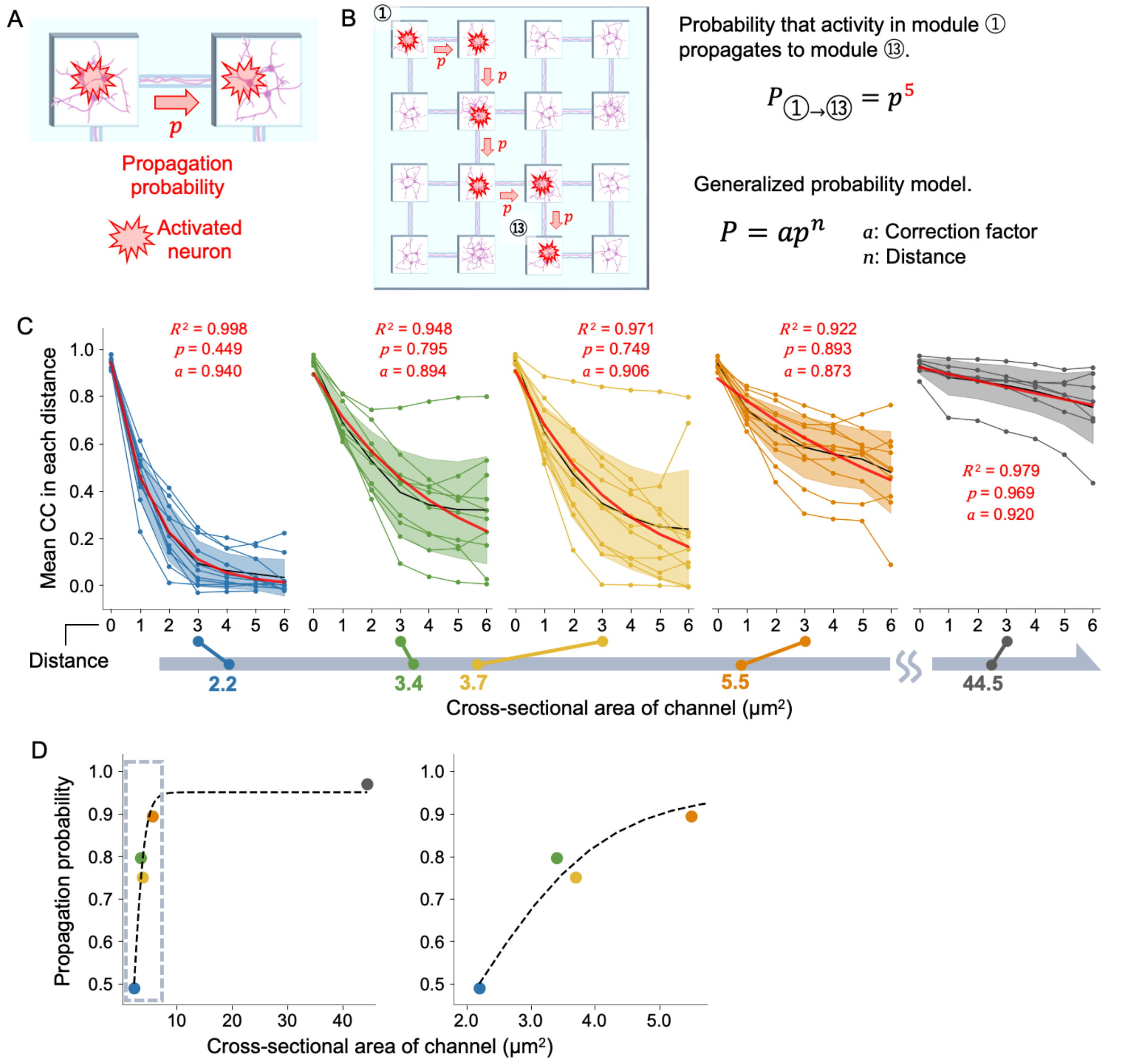

**Figure S3.** Probabilistic model describing the relationship between correlation coefficients (CCs) and distance. (A) The propagation probability $p$ is the probability of the activity of a module propagating to neighboring modules. (B) Example of propagation probability at distance 5 and the generalized probability model. (C) The propagation probability model (red lines) fits the mean curve of CC (black lines) with high accuracy. CC against distance in each network (colored lines). The mean value at each distance is plotted (black lines), with shading indicating the standard deviation (SD). (D) Relationship between the cross-sectional area of the channels and the propagation probability. A sigmoidal curve fit is shown to aid visualization (dashed line). Right: a magnified image of the area surrounded by the dashed grey line in the left panel.

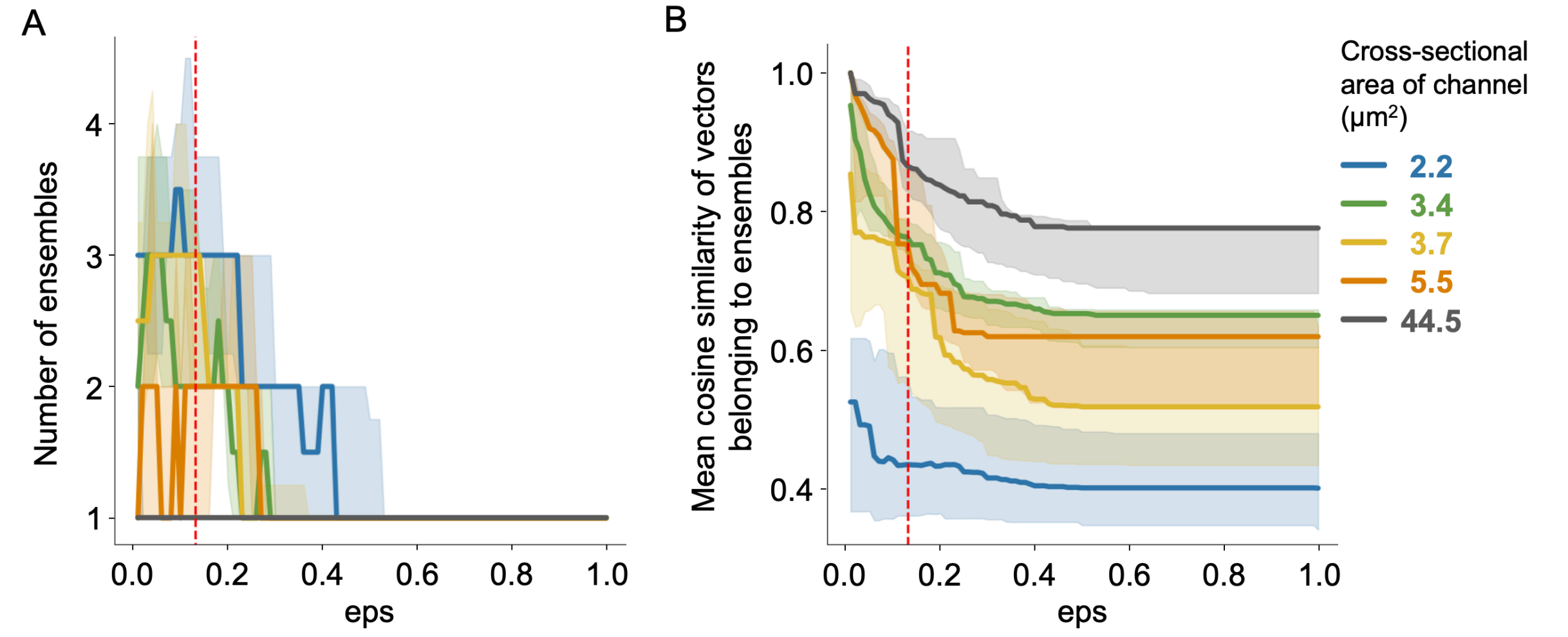

**Figure S4.** Influence of hyperparameter 'eps' (epsilon) in DBSCAN clustering. Each line denotes the median of the networks, and the filled area indicates between the first and third quantiles. The dashed red line indicates eps = 0.13 in this study. (A) Relationship between epsilon and the number of ensembles. (B) Relationship between epsilon and the mean cosine similarity of population vectors belonging to ensembles.

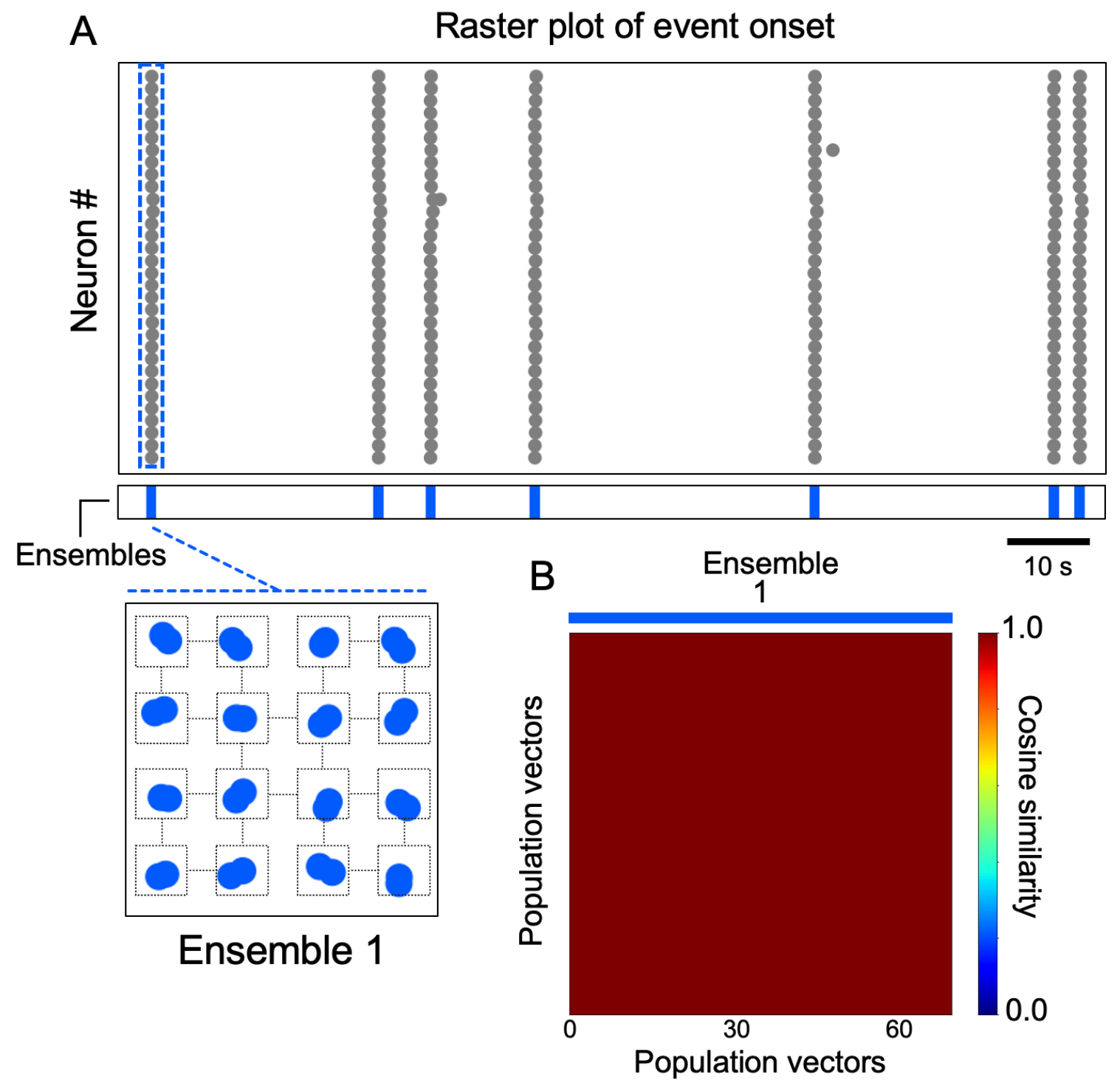

**Figure S5.** Neuronal ensembles in the wide-channel network (channel size: 44.5 μm$^2$). (A) Raster plot and detected neuronal ensembles. Top: raster plot of event onset and detected neuronal ensembles corresponding to the raster plot. Bottom: the spatial map of the neurons belonging to the ensemble. (B) Cosine similarity matrix of population vectors belonging to neuronal ensembles in the representative network. The colored line above the similarity matrix indicates the range of population vectors corresponding to each ensemble.

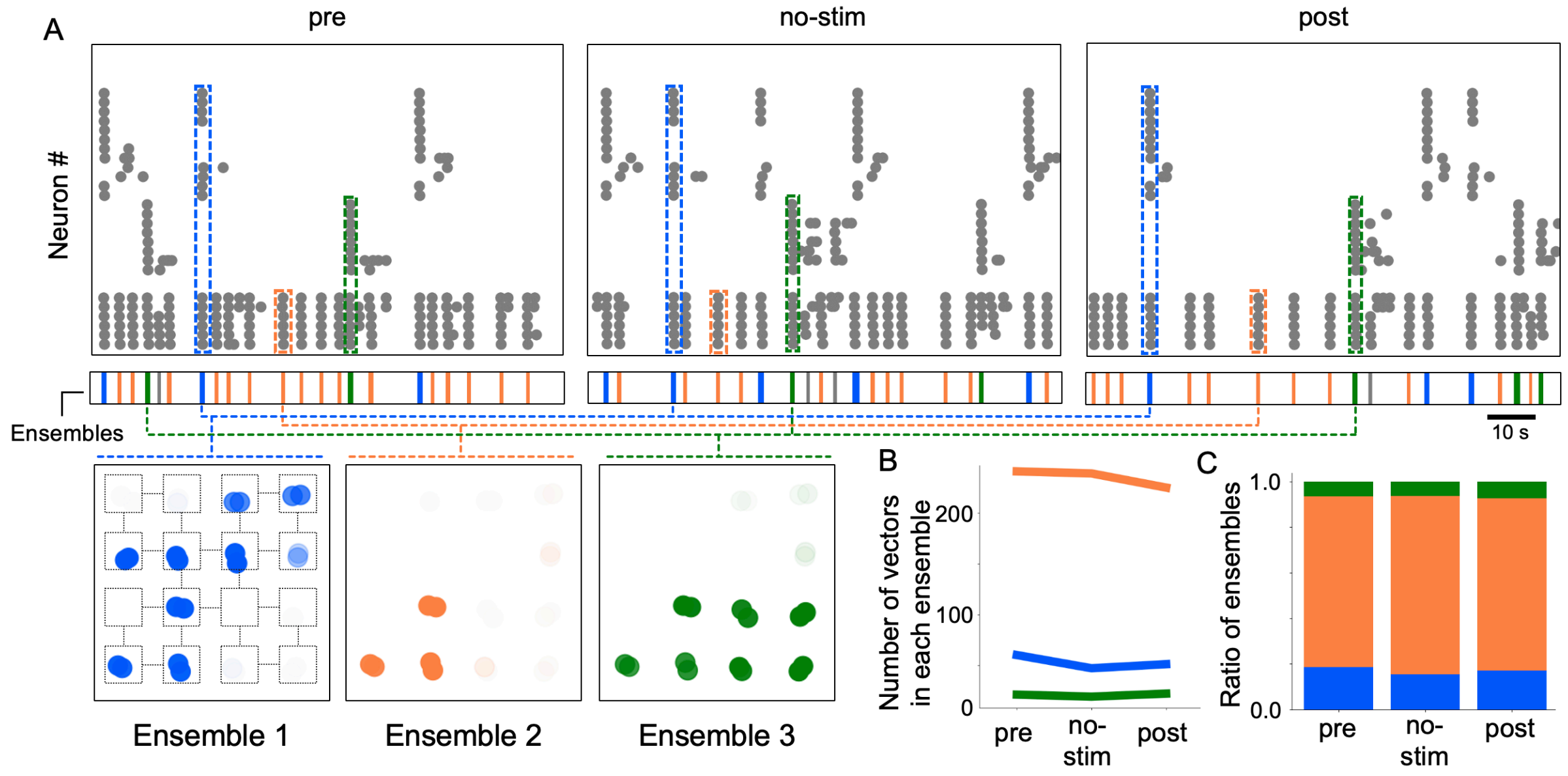

**Figure S6.** Neuronal ensembles were stable in the absence of stimulation in the shallow-channel network (channel size: 3.7 μm$^2$). (A) Raster plot of event onset and detected neuronal ensembles in a representative network. Top left: prestimulation session. Top middle: nonstimulation session. Top right: poststimulation session. Bottom: Spatial map of neurons belonging to neuronal ensembles. The gradation of a neuron indicates the contribution ratio to the ensemble. (B) Number of population vectors belonging to ensembles in each session of the network. (C) Proportion of population vectors belonging to ensembles in each session of the network.